\documentclass[aps,superscriptaddress,preprint,nofootinbib]{revtex4}

\usepackage{amsmath}
\usepackage{graphicx}
\usepackage{amssymb}
\usepackage{dcolumn}
\usepackage{bm}
\usepackage{color}

\makeatletter

\usepackage{epstopdf, epsfig}
\usepackage{esint}
\DeclareMathAlphabet\mathbfcal{OMS}{cmsy}{b}{n}

\begin{document}

\title[R. Maity and P.S. Burada]{Unsteady chiral swimmer and its response to a chemical gradient}

\author{Ruma Maity}\thanks{rumamaity@phy.iitkgp.ac.in} 
\affiliation{Department of Physics, Indian Institute of Technology Kharagpur, Kharagpur 721302, India}

\author{P. S. Burada}\thanks{Corresponding author: psburada@phy.iitkgp.ac.in}
\affiliation{Department of Physics, Indian Institute of Technology Kharagpur, Kharagpur 721302, India}


\begin{abstract}
Unsteadiness occurs in the motion of swimmers while they start from rest or escape from a predator, or attack prey. In this paper, we study the behavior of an unsteady chiral swimmer, with a prescribed surface slip velocity, in the low Reynolds number regime and its response to an external chemical gradient. In the first part, by solving
the unsteady Stokes equation, we calculate the migration velocity $(\bf{U})$, the rotation rate $(\bm{\Omega})$, and the flow field of the unsteady swimmer in a closed-form. We compare these results with some previously known results in appropriate limits. In the second part, we investigate the response of the unsteady chiral swimmer to an external chemical gradient, which can influence the swimmer's surface slip velocity. 
Consequently, the swimmer either steers towards the source of the chemical gradient or moves away from it, depending on the strength of $\bf{U}$ and $\bm{\Omega}$ and the corresponding angle $(\chi)$ between them. Interestingly, the swimmer swims in a closed orbit near the chemical target, depending on the strengths of $\Omega$ and $\chi$. This diffusive orbit is nothing but the transition state between the successful and unsuccessful chemotaxis. We present a complete state diagram representing the successful, unsuccessful, and orbiting states for various strengths of $\Omega$ and $\chi$. This study is useful to understand the unsteady propulsion of ciliated microorganisms and their response to external gradients. 
\end{abstract}

\maketitle

\section{Introduction}

\textit{Volvox, Marine Zooplankton, Paramecium}, Sperm cell, {\it E. Coli}, etc., are the low Reynolds number microswimmers. They do not need any external force to swim in the fluid. While some use flagella and cilia, others modulate their body shape asymmetrically to move in the fluid. 
It has been found that the microorganisms can use different swimming gaits to escape from the adverse environment \citep{hamel}. 
Some of them exhibit only translational motion \citep{lighthill, blake}, while the others swim in helical paths \citep{friedrichth}, having both translational and rotational movements. 
In general, microorganisms generate unsteady flows while swimming due to the sudden start from rest in a still medium. 
Planktonic organisms, which significantly affect trophic dynamics, are the most common examples of unsteady swimmers. Some of them exhibit oscillatory flows due to the synchronous movement of the appendage attached to the surface 
of the body \citep{klindt}. Guasto \textit{et al.} \citep{guasto} have experimentally measured the time-dependent velocity field of \textit{C. reinhardtii}. 
Unsteadiness can also occurs due to several other reasons, for example, $(i)$ the unsteady beating of the appendages attached to the surface of the body, $(ii)$ velocity fluctuations developed in the presence of the predator and prey, and $(iii)$ turbulent ambient flow \citep{magar, crawford}.

The time-dependent movement of a passive body is well studied \citep{basset, auton, gprs}. 
However, the unsteady motion of microorganisms is not explored much in the past. 
The squirmer model \citep{lighthill, blake} was extended to study the unsteady motion of the ciliated microorganisms by Rao \citep{rao}.
Later, this model was extended by Ishimoto \textit{et al.} to study the motion of an unsteady inertial squirmer with a small surface deformation under the action of gravity \citep{ishimoto}. Recently, Magar \textit{et al.} used the squirmer model \citep{lighthill, blake}, which exploits time-dependent but quasi-steady flow fields around the body to explain the nutrient uptake of the unsteady but inertia-free squirmer \citep{magar}. 
Moreover, as an unsteady squirmer swims in a nutrient-enriched environment, its nutrient uptake rate increases with increasing speed \citep{magar}. 
On the other hand, Wang \textit{et al.} \citep{wang} derived the equation of motion of an unsteady inertial squirmer propelling in a background of a non-uniform flow field. 
Notably, the biogenic mixing in the environment is aided by a collection of unsteady microswimmers \citep{mueller, thiffeault, lin}.  Unsteadiness also results in the microorganisms' motion due to the entrapment of the body at the water-air interface, as experimentally observed for \textit{Tetrahymena} \citep{ferracci}. 
Note that the scallop theorem is also invalid for unsteady squirmer if the inertial effects are taken into account \citep{wang, lauga}. 

 In unsteady swimming, while the dominating viscous force keeps the Reynolds number $(Re = \rho U a/\eta)$ low, the unsteady inertia enters through the Strouhal number $(Sl)$. The Strouhal number is defined as $Sl = a \omega/U$, where
$a/U$ is the convective timescale and 
$\omega$ is the frequency corresponding to the oscillatory flow of the swimmer. 
Consequently, $SlRe$ measures the relative magnitude of unsteady inertia. Note that $a$ is the characteristic length-scale of the body, $U$ is the characteristic velocity of the body, $\rho$ is the fluid density, and $\eta$ is the fluid viscosity. Thus, for a higher $Sl$, the oscillation in the flow, i.e., unsteady inertia, dominates.  Moreover, unsteady inertia dominates the hydrodynamic forces while the vorticity around the swimmer has not been diffused out to a large distance \citep{lovalenti}. Notably, the convective inertia is negligible in the previous case. We assume $ Re \ll Sl$ in this work. The former assumptions work well for many microorganisms. For example, for an algal cell $Re = 0.001$, and $Sl = 1$ \citep{guasto}. The unsteady inertial forces acting on the unsteady swimmer have the following terms-Basset force and added mass force \citep{basset}.  The Basset or history force is due to the lagging boundary layer around the unsteady swimmer. 
On the other hand, the added mass force arises as the body accelerating in a medium generates an instantaneous pressure field in the surrounding medium. 
Hence, not only the body but a portion of the medium accelerates with the body giving rise to added mass effect.

In biological systems, though diffusion influences many transport phenomena, directed motion is also significant. 
The term \textit{taxis} refers to the locomotion of the cell in response to its environment \citep{houten,mitchell}. 
If the external stimulus is a chemical substance, then the microorganism's response by changing its course of motion is known as chemotaxis. If the microorganism gets attracted by the stimulus, it is called positive chemotaxis, and if repelled, it is called negative chemotaxis \citep{tsang}. Both natural and artificial swimmers can exhibit chemotaxis \citep{jin, hong, paxton, geiseler}. 
For example, artificial swimmers like Janus particles are driven by self-thermophoresis created by laser irradiation \citep{jiang}, nanorobots \citep{sengupta} can accelerate subjected to external stimuli. 
Experimentally it has been found that designed magnetic micro-swimmers can be used as an effective transport agent \citep{kokot} for drug delivery.  
Also, the confined motion of the magnetic colloidal particles in the presence of a sinusoidal magnetic field is useful for designing microfluidic devices \citep{tierno}. The time-dependent magnetic field can profoundly influence the translational-rotational coupling in the motion of the artificial swimmers giving rise to different types of swimming paths, for example, helical and superhelical.
Bio-inspired unsteady artificial swimmers have biomedical applications \citep{ceylan}.
Though the accelerated motion of microorganisms is energetically expensive, it is the only way out for them under the attack of a predator. 
Therefore, it is crucial to understand the unsteady bio-locomotion to gain better knowledge about various natural phenomena and biomedical applications involved with them.

In this article, we aim to study the hydrodynamic behavior of an unsteady chiral swimmer (non-axisymmetric). 
We use the unsteady version of the generalized squirmer model called the chiral squirmer \citep{burada,maity}, a non-deformable spherical body free from external forces and torques. 
However, the swimmer experiences forces from its own generated flow field. 
We solve the corresponding unsteady Stokes equation with appropriate boundary conditions, using the double curl representation \citep{padmavathi}. Note that the former method has been proved to provide a complete, general solution to the unsteady Stokes equation \citep{venkatalaxmi}. We provide the flow field, migration velocity, and rotation rate of an unsteady chiral swimmer in a closed-form. 
The migration velocity has been compared with the known results in the axisymmetric limit. Later, we numerically investigate the response of the unsteady chiral swimmer to an external chemical stimulus. 
The paper is organized as follows. In \S~\ref{sec:model}, we introduce the unsteady chiral squirmer model. In \S~\ref{comp}, we discuss the hydrodynamic flow fields of the swimmer. \S~\ref{sec:chem_grad} provides the response of an unsteady swimmer to an external chemical gradient. The main conclusions are provided in \S~\ref{sec:conclusions}.

\section{Unsteady chiral swimmer model}
\label{sec:model}

 In general, when microswimmers swim in a fluid, the inertial effects can usually be neglected as they belong to the low Reynolds number regime. However, unsteady or oscillatory swimming may be important in many biological phenomena. The hydrodynamic flow field generated by a swimmer then obeys the unsteady Stokes equation \citep{Happel},
\begin{align}
  \rho \frac{\partial \mathbf{v}}{\partial t} 
   = - \nabla p + \eta\nabla^2 \mathbf{v} \ , \quad
  \nabla \cdot \mathbf{v} = 0 \ ,
\label{us1}
\end{align}
where $\mathbf{v}$ and $p$ are the velocity and pressure fields, respectively.
$\rho$ and $\eta$ are the density and viscosity of the surrounding fluid, respectively. 
The non-dimensionalization of Eqs.~(\ref{us1}) is provided in appendix \ref{appendix_stre}. 
Note that hereafter we use the dimensionless description. 
There are minimal works in literature determining the general solution of the unsteady Stokes equation \citep{rao, padmavathi}. In this study, we have considered the solution by Venkatalaxmi \textit{et al.} \citep{padmavathi} to determine the flow field around a chiral swimmer. The details are provided in the appendix \ref{appendix}. 

\begin{figure}
\centering
\includegraphics[scale=0.14]{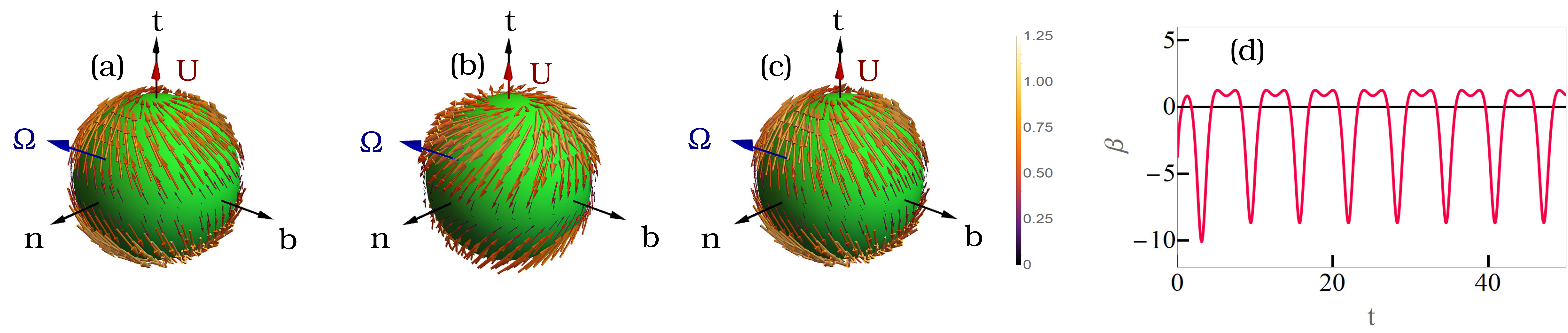}
\caption{(Color Online)  
Real part of the slip velocity (Eq.~\ref{slip}) 
of an unsteady chiral swimmer at different dimensionless 
times $t = 0 (a)$, $t = 26 (b)$ and $t = 28 (c)$. 
The swimmer propels with a time dependent velocity $\bm{U}$ and rotation rate $\bm{\Omega}$. 
$(\mathbf{n},\mathbf{b},\mathbf{t})$ is the body frame of reference with origin 
at the center of the body. 
The color bar depicts the strength of the surface slip velocity. 
The parameter values are set to, 
$\delta_{10}^{0\, A} = 2.39$, $\delta_{10}^{1\, A} = 1.4$,
$\delta_{10}^{2\, A} = -0.09$, $\delta_{10}^{\prime 1\, A} = 2.6$, $\delta_{10}^{\prime 2\, A} = -0.16,$
$\delta_{20}^{0\, A} = 0$, $\delta_{20}^{1\, A} = 5.457$, $\delta_{20}^{2\, A} = -2.353$, 
$\delta_{20}^{\prime 1\, A} = -5.457$, $\delta_{20}^{\prime 2 \, A} = -2.353$, 
$\xi_{20}^{0\, C} = 2,\xi_{20}^{1 \, C} = 2,
\xi_{20}^{2\, C} = 3,\xi_{20}^{\prime 1\, C} = -2,\xi_{20}^{\prime 2\, C} = -3$.
Note that $n = 1, m = 0,1$ modes of the rotational part of the flow field do not contribute to the hydrodynamic field in the lab frame of reference. $(d)$ The time-dependent stresslet $(\beta)$ has been plotted as a function of time $(t)$ depicting the changing nature of the swimmer with time. Here, $\beta>0$ is a puller, $\beta<0$ is a pusher, $\beta=0$ is a neutral swimmer.}
\label{fig:slip_vel}
\end{figure}

A chiral squirmer is a rigid spherical body with a tangential active surface slip.
However, in the current study, the active slip $\mathbfcal{V}_s (\theta, \phi, t)$ is time-dependent.
It is parameterized by the polar and azimuthal angles $\theta$ and $\phi$, respectively,  and defined 
in a body-fixed frame $(\bf{n}, \bf{b}, \bf{t})$ in terms of gradients of spherical harmonics that form a basis for tangential vectors on the surface. 
The general unsteady slip on the surface of the spherical body is prescribed as,
\begin{align}
\mathbfcal{V}_s (\theta, \phi, t) &= \mathbf{v}_s (\theta, \phi, t) + \mathbf{v}_s^d (\theta, \phi, t) \, , 
\label{slip}
\end{align}
where,
\begin{align}
\mathbf{v}_s (\theta, \phi, t)  & = 
 \sum\limits_{l=1}^\infty\,\sum\limits_{m= 0}^l\, \sum\limits_{j=0}^\infty \,
\Big[-\delta_{l m }^j \nabla_s S_{l m}^j(\theta, \phi) 
 + \xi_{l m}^j 
\mathbf{e_r} \times \nabla_s T_{l m}^j(\theta, \phi)\Big] e^{\lambda_j^2 t} \, , 
\label{slip_s}
\\
\mathbf{v}_s^d (\theta, \phi, t) & = 
\sum\limits_{l=1}^\infty\,\sum\limits_{m= 0}^l\,\sum\limits_{k=1}^\infty \, 
\Big[-\delta_{l m }^{\prime k} \nabla_s S_{l m}^{\prime k}(\theta, \phi) + \xi_{l m}^{\prime k} 
\mathbf{e_r} \times \nabla_s T_{l m}^{\prime k}(\theta, \phi)\Big] e^{\lambda_k^2 t} \,,
\label{slip_ex}
\end{align}
where $\delta_{l m}^j$, $\xi_{l m}^j$, $\delta_{l m}^{\prime k}$ and $\xi_{l m}^{\prime k}$ are the slip coefficients.
$\nabla_s$ = $\mathbf{e_\theta} \frac{\partial}{\partial \theta} + \mathbf{e_ \phi} \frac{1}{\sin \theta} \frac{\partial}{\partial \phi}$ is the surface gradient operator. 
$S_{l m}^j(\theta, \phi)$ and $T_{l m}^j(\theta, \phi)$ 
are the spherical harmonics having the form, 
$S_{l m}^j(\theta, \phi) = P_l^m(\cos \theta) (A_{l m}^j \cos m\phi + B_{l m}^j \sin m\phi)$ and 
$T_{l m}^j(\theta, \phi) = P_l^m(\cos \theta) (C_{l m}^j \cos m\phi + D_{l m}^j \sin m\phi)$, respectively. $P_l^m (\cos \theta)$ are the associated Legendre polynomials. 
Similarly, 
$S_{l m}^{\prime k}(\theta, \phi) = P_l^m(\cos \theta) (A_{l m}^{\prime k} \cos m\phi + B_{l m}^{\prime k} \sin m\phi)$ 
and 
$T_{l m}^{\prime k}(\theta, \phi) = P_l^m(\cos \theta) (C_{l m}^{\prime k} \cos m\phi + D_{l m}^{\prime k} \sin m\phi)$.
 Note that the slip velocity is decomposed into two parts. 
The first part $(\mathbf{v}_s)$ corresponds to the oscillatory part and the second one $(\mathbf{v}_s^d)$ corresponds to the decaying part consistent with the general solution of the unsteady Stokes equation. Also, while $\mathbf{v}_s$ is essential to capture the long-time unsteady behaviour of the swimmer, $\mathbf{v}_s^d$ is necessary to capture the transient motion of the swimmer.

We use the above slip velocity and calculate the flow field generated by the unsteady chiral swimmer in the laboratory frame of reference (see the Eqs.~\ref{eq:gen_comp}). 
In the flow field solution, $\lambda^2_{n}\, (n = j, k)$ can be either complex or real. Therefore, to deal with the complete solution, we consider both $\lambda^2_j = i \,j\, $ and $\lambda_k^2 = - k\, $. Here, $i = \sqrt{-1}$. We can see that $\lambda_{j}^2$ is a constant related to the frequency of the body's oscillatory motion, while $\lambda_{k}^2$ belongs to its transient motion. Note that $j = k = 0$ correspond to the steady part of the slip. However, to avoid repetition, we consider the case $k > 0$ in the current study. 
 Note that, like in the case of a steady swimmer, while the first mode in the slip velocity 
$(l = 1)$ is related to the translational and rotational motion of the swimmer, $l = 2$ mode corresponds to the leading order flow field. In general, the contribution of the higher-order modes $(l > 2)$ can be ignored. Thus, we include only $l = 1, 2$ modes in the surface slip for the rest of the article.
It may be noted that in the limit $\lambda_{j,k} \rightarrow 0$, by setting 
$\xi_{l m}^{j,\prime k} = 0$ in the surface slip (\ref{slip}), one can retrieve the simple axisymmetric squirmer model introduced by Lighthill \citep{lighthill}. However, in the same limit, by setting only $\xi_{l m}^{j, \prime k} \neq 0$, the chiral squirmer model can be retrieved \citep{burada}. 
Thus, the surface slip (\ref{slip}) in the current model is very general. 
In the following, for simplicity, we consider only the first few oscillatory modes of the surface slip, i.e., $j = 0,1,2$ and $k = 1, 2$, and neglect higher frequency modes. 
This time-dependent slip causes the swimmer to change its nature cyclically with time from puller to pusher and vice versa, see Fig.~\ref{fig:slip_vel}. 
 The nature of the swimmer can be captured through the parameter $\beta$, see Fig.~\ref{fig:slip_vel}(d), which is defined as
\begin{align}
\beta &= \frac{\sum\limits_{m= 0}^2\,\sum\limits_{k=1}^\infty \, \sum\limits_{j=0}^\infty \,
\delta_{2 m }^{j, \prime k} e^{\lambda_{j,k}^2 t} }{\sum\limits_{m= 0}^1\,\sum\limits_{k=1}^\infty \, \sum\limits_{j=0}^\infty \, 
\delta_{1 m }^{j, \prime k}e^{\lambda_{j,k}^2 t} }\, .
\end{align}
Here, we considered only $j = 0, 1, 2$ and $k = 1,2$. 
Consequently, the swimmer changes its nature depending on its slip coefficients, see Fig.~\ref{fig:slip_vel}(d).  
Note that the former behavior is common in many microorganisms \citep{klindt, wu, hintsche}. 
However, a steady squirmer never changes its nature of swimming over time. 
The instantaneous surface velocity determines the swimming speed and the rotation rate of the unsteady swimmer. 

\subsection{Velocity and rotation rate}
\label{sec:Vel-rot}

\begin{figure}
\centering
\includegraphics[scale=0.16]{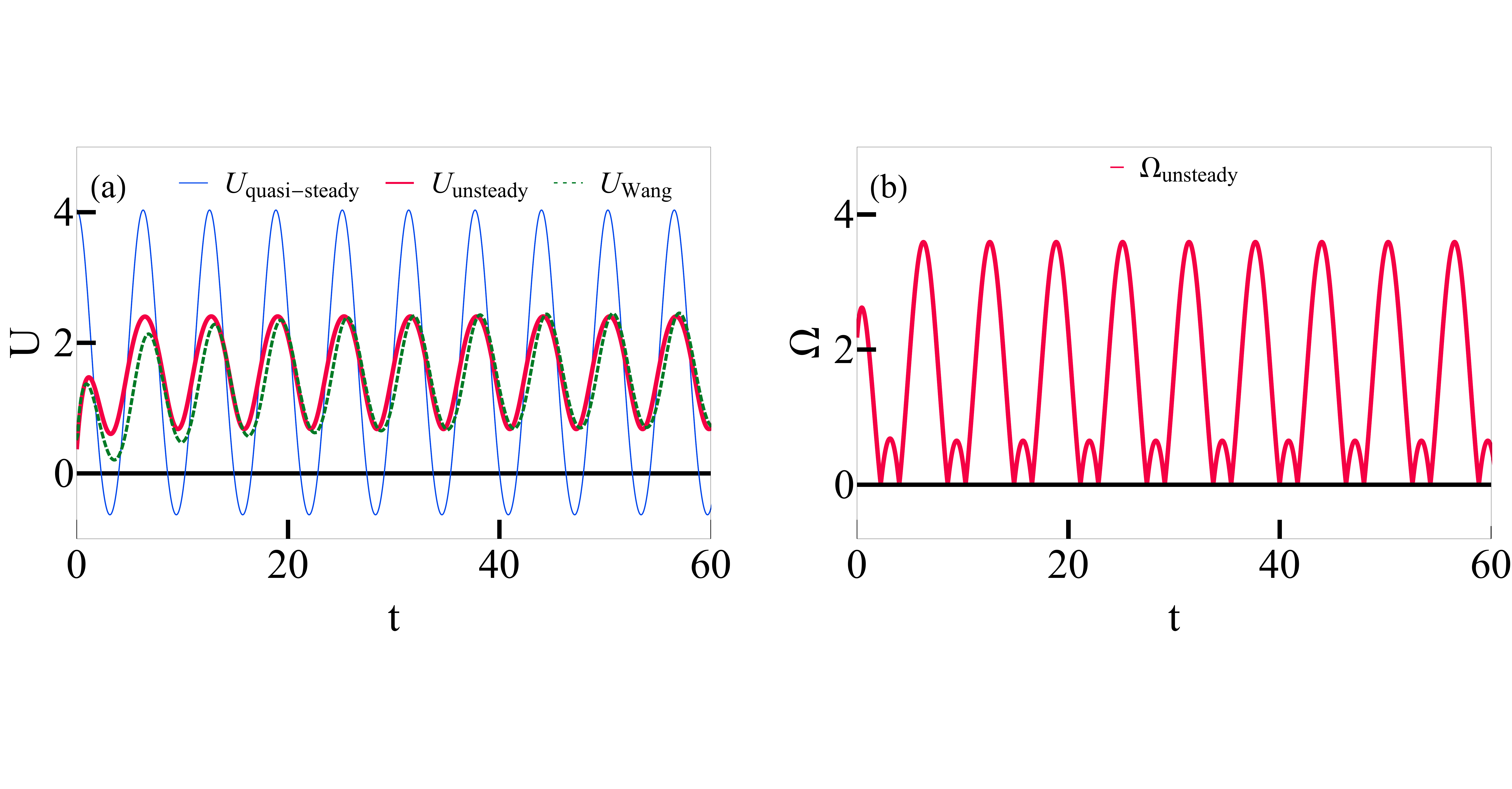}
\caption{(Color Online) 
$(a)$ The dimensionless swimming velocity $U$ of the unsteady chiral swimmer is compared with the results of Wang \textit{et al.} and Blake \textit{et al}.
All the corresponding parameters have the same values as in the Fig.~\ref{fig:slip_vel}. The velocity by Blake, $U_{\text{quasi-steady}}$, and Wang, $U_{\text{Wang}}$, are plotted with $B_{10} = 2.39, B_{11} = 3.5, B_{12} = 0.16$ and $SlRe = 10$ \citep{wang}. $(b)$ Rotation rate $(\Omega)$ of the chiral swimmer is plotted as a function of time with $\xi_{10}^{0\, C}= \xi_{11}^{0\, C} = 1,\xi_{10}^{1\, C}= \xi_{11}^{1\, C} = 1.5,\xi_{10}^{2\, C}=\xi_{11}^{2\, C} = 0.04,\xi_{10}^{\prime 1\, C}= \xi_{11}^{\prime 1\, C} = -0.5,\xi_{10}^{\prime 2\, C}= \xi_{11}^{\prime 2\, C} = -0.5$. Here, $\Omega = \sqrt{\Omega_n^2 + \Omega_t^2}$ where 
$\Omega_n$ and $\Omega_t$ are the components of the rotation rate along $\mathbf{n}$ and $\mathbf{t}$ axes, respectively.}
\label{fig:vel_om_compare}
\end{figure}

Using the Lorentz reciprocal theorem \citep{kim} for the unsteady Stokes flow, 
we calculate the velocity and rotation rate of the swimmer. 
The general expressions are given by (see appendix \ref{appendix2}),
\begin{align}
\label{eq:U}
\mathbf{U}(t) & = \mathbf{U}_0 
+ \sum_{j=1}^{\infty}  \Re \left[ \mathbf{U}^j \, e^{\lambda^2_j t} \right] + \sum_{k=1}^{\infty}  \Re \left[ \mathbf{U}^{\prime k} \, e^{\lambda^2_k t} \right]\,,\\
\label{eq:W}
\bm{\Omega}(t) & = \bm{\Omega}_0 + 
\sum_{j=1}^{\infty}  \Re \left[ \bm{\Omega}^j \, e^{\lambda^2_j  t} \right] + 
\sum_{k=1}^{\infty}  \Re \left[ \bm{\Omega}^{\prime k} \, e^{\lambda^2_k t} \right]\,,
\end{align}
where  
\begin{align}
 \label{eq:U_0}
 \mathbf{U}_0 & = \frac{2}{3}\, \left(\delta_{1 1}^{0A}\,\mathbf{n} + \delta_{1 1}^{0B}\,\mathbf{b} + \delta_{1 0}^{0 A}\,\mathbf{t}\right) \,,\\
 \label{eq:U_j}
\mathbf{U}^{j, \prime k} & = \frac{2}{3} \,
\frac{(1+ a \lambda_{j,k})}{\left(1 + a \lambda_{j,k} + \frac{a^2 \lambda_{j,k}^2}{3}\right)} \, \left(\delta_{1 1}^{j, \prime k \, A}\,\mathbf{n} + \delta_{1 1}^{j,\prime k \, B}\,\mathbf{b}+ \delta^{j,\prime k \, A}_{1 0}\,\mathbf{t}\right)\,, \\
 \label{eq:Omega_0}
\bm{\Omega}_0 & = \frac{\xi_{1 1}^{0C}}{a} \mathbf{n} + \frac{\xi_{1 1}^{0D}}{a}  \mathbf{b} +  \frac{\xi^{0 C}_{1 0}}{a} \mathbf{t} \, ,\\
 \label{eq:Omega_j}
\bm{\Omega}^{j, \prime k} & = \frac{\xi_{1 1}^{j,\prime k \, C}}{a} \mathbf{n} + \frac{\xi_{1 1}^{j,\prime k \, D}}{a}  \mathbf{b} +  \frac{\xi^{j,\prime k \, C}_{1 0}}{a} \mathbf{t}  \,.
\end{align}
Here, 
$\delta_{1 0}^{j,\prime k \, A} = \delta_{1 0}^{j,\prime k} \cdot A_{ 1 0}^{j,\prime k}$,
$\xi_{1 0}^{j,\prime k \,C} = \xi_{1 0}^{j,\prime k} \cdot C_{ 1 0}^{j,\prime k}$,
$\xi_{1 1}^{j,\prime k \, C} = \xi_{1 1}^{j,\prime k} \cdot C_{ 1 1}^{j,\prime k}$,
$\xi_{1 1}^{j,\prime k \, D} = \xi_{1 1}^{j,\prime k} \cdot D_{ 1 1}^{j,\prime k}$,
and $A_{lm}^{j,\prime k}, B_{lm}^{j,\prime k}, C_{lm}^{j,\prime k}$ and $D_{lm}^{j,\prime k}$ are the coefficients of $S_{l m}^{j, \prime k}$ and 
$T_{l m}^{j, \prime k}\, (j = 0,1\, \text{and}\, k = 1)$.
In the rest of the article, we set the parameters 
$\delta^{0\, A}_{1 1} = \delta^{0\, B}_{1 1} = \xi^{0 \, D}_{1 1} = 0$, 
and also 
$\delta^{j, \prime k \, A}_{1 1} = 
\delta^{j, \prime k \, B}_{1 1} =  \xi^{j,\prime k \, D}_{1 1} = 0$, for any $j$,  
(in eqs.~\ref{eq:U_0}, \ref{eq:U_j}, \ref{eq:Omega_0}, and \ref{eq:Omega_j})
such that the body translates along $\mathbf{t}$ direction, i.e. $\bm{U}(t) = U_t \,\bm{t}$ (see Fig.~\ref{fig:slip_vel}) and having the rotation rate in $\mathbf{n, t}$ plane. 

Fig.~\ref{fig:vel_om_compare} depicts the velocity and rotation rate of the unsteady chiral swimmer as a function of time. 
Surprisingly, the velocity obtained in our calculations is very close to the one calculated by Wang \citep{wang} (see Fig.~\ref{fig:vel_om_compare}a). While the unsteady velocity \citep{wang}  counts the contribution of the added mass and the Basset force, velocity in the quasi-steady limit considers only the hydrodynamic force \citep{blake, magar}. Wang and Ardekani \citep{wang} analytically derived the fundamental equation of motion of unsteady low-Reynolds-number propulsion in an unbounded medium without determining the hydrodynamic flow field around the body. However, the determined unsteady migration velocity of the body \citep{wang} is a function of the product of the \textit{Strouhal} and \textit{Reynolds} numbers. Conversely, the unsteady velocity determined in our case purely depends on the coefficients of the prescribed time-varying surface slip velocity of the squirmer. 
 The $Sl Re$ is absorbed into the slip coefficients of the surface slip velocity in the former case (see appendix \ref{appendix_stre}).
Additionally, the body may possess a time-dependent rotation rate, as we have plotted in Fig.~\ref{fig:vel_om_compare}(b). 
Here, we have considered unsteady rotational motion having components along the $\mathbf{t}$ and $\mathbf{n}$-axes. 
The rotation rate is oscillatory, and the behavior is determined by 
$\Omega = \sqrt{\Omega_n^2 + \Omega_t^2}$. 
Due to this choice, the unsteady swimmer moves in super helical paths 
because $\Omega_n$ and $\Omega_t$ are time dependent. 
If they had been time-independent, we would have obtained a helical path \citep{burada}.
Note that the added mass and the Basset force do not influence the magnitude of $\bm{\Omega}(t)$, see eqs.~\ref{eq:Omega_0} \& ~\ref{eq:Omega_j}.

Solving the force- and torque balance equations for the chiral swimmer,
the equations of motion of the swimmer can be obtained, which are analogous to 
Frenet-Serret equations. They read, 
\begin{align}
\label{eqn:single}
\dot{\mathbf{r}} & = \mathbf{U}(t), \nonumber \\
\dot{\mathbf{n}} & = \bm{\Omega}(t)\times \mathbf{n},
\quad 
\dot{\mathbf{b}} = \bm{\Omega}(t)\times \mathbf{b},
\quad 
\dot{\mathbf{t}} = \bm{\Omega}(t)\times \mathbf{t}\,.
\end{align}
The swimming path, generated by Eqs.~(\ref{eqn:single}), depends 
on the strengths of $\mathbf{U}(t)$ and $\bm \Omega (t)$, and the coupling between them. 
In the case of a steady chiral swimmer, the path of the swimmer is a helix \citep{burada}. 
However, in the present case, the resultant swimming path is not helical but superhelical or deformed helical (see Fig.~\ref{traj_rad}(b)). 
Notably, a similar superhelical path has been observed for flagellated microorganisms modeled with the three-sphere model \citep{cortese}.

\begin{figure}
\centering
\includegraphics[width=1\textwidth]{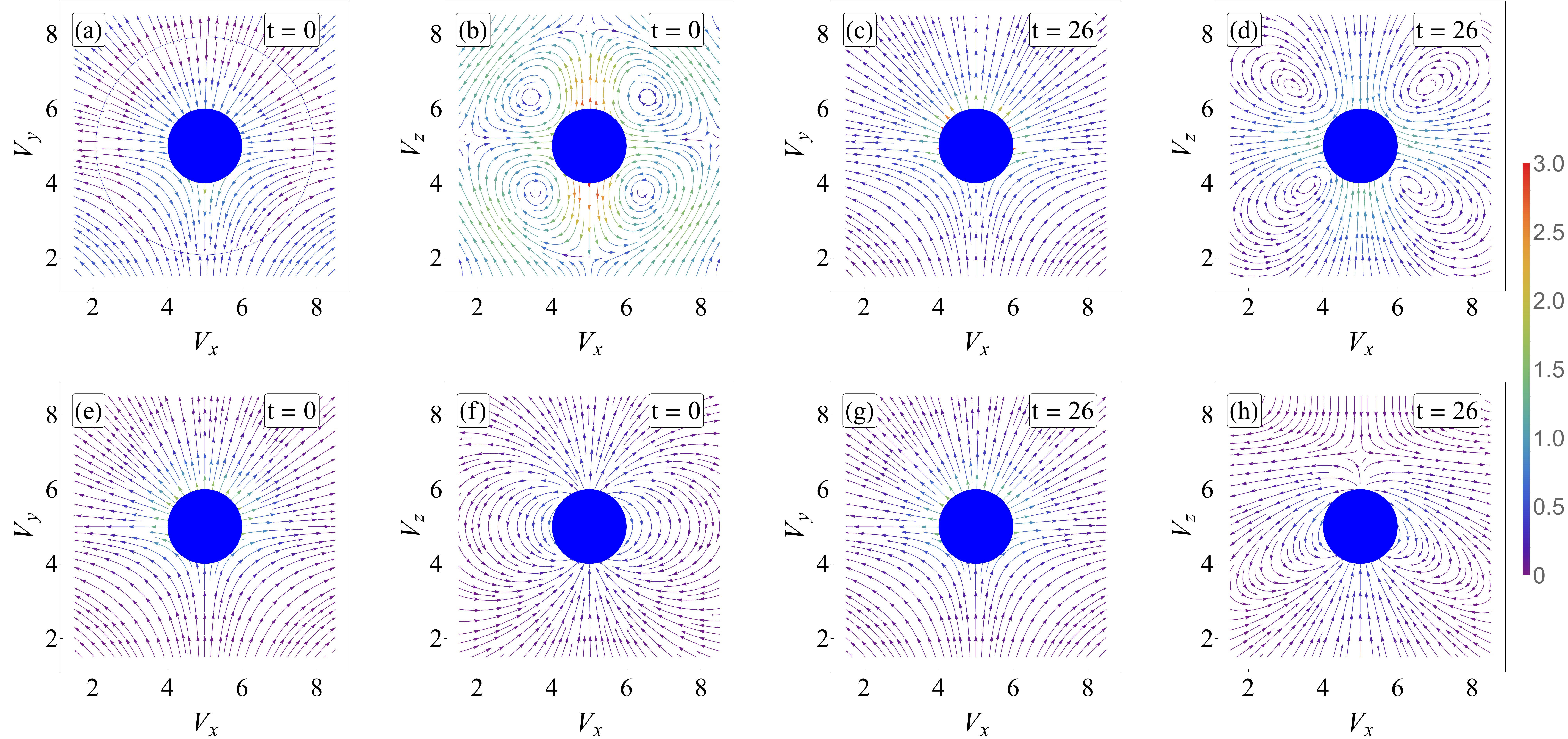}
\caption{(Color online) 
Real part of the flow field around an axisymmetric swimmer in different planes at different times.
$(a)-(d)$ are the flow fields of unsteady swimmer, in the absence of chirality, 
compared with that of a quasi-steady swimmer \citep{magar} $((e)-(h))$. 
(a), (b), (e), (f) at $t = 0$ and (c), (d), (g), (h) at $t = 26$.
For the unsteady swimmer parameters are used as in Fig.~\ref{fig:slip_vel}. 
For the quasi-steady swimmer, the parameters are 
$B_{10} = 2.39, B_{11} = 3.5, B_{12} = 0.16, B_{20}=0, B_{21} = 5.457$ and $B_{22} = -2.353$ \citep{magar,wang}.
The stagnation regions are the common feature of unsteady swimmers, shown by a circle in $(a)$. 
The color bar indicates the strength of the velocity field.}
\label{vel_stream}
\end{figure}
\begin{figure}
\centering
\includegraphics[width=0.5\textwidth]{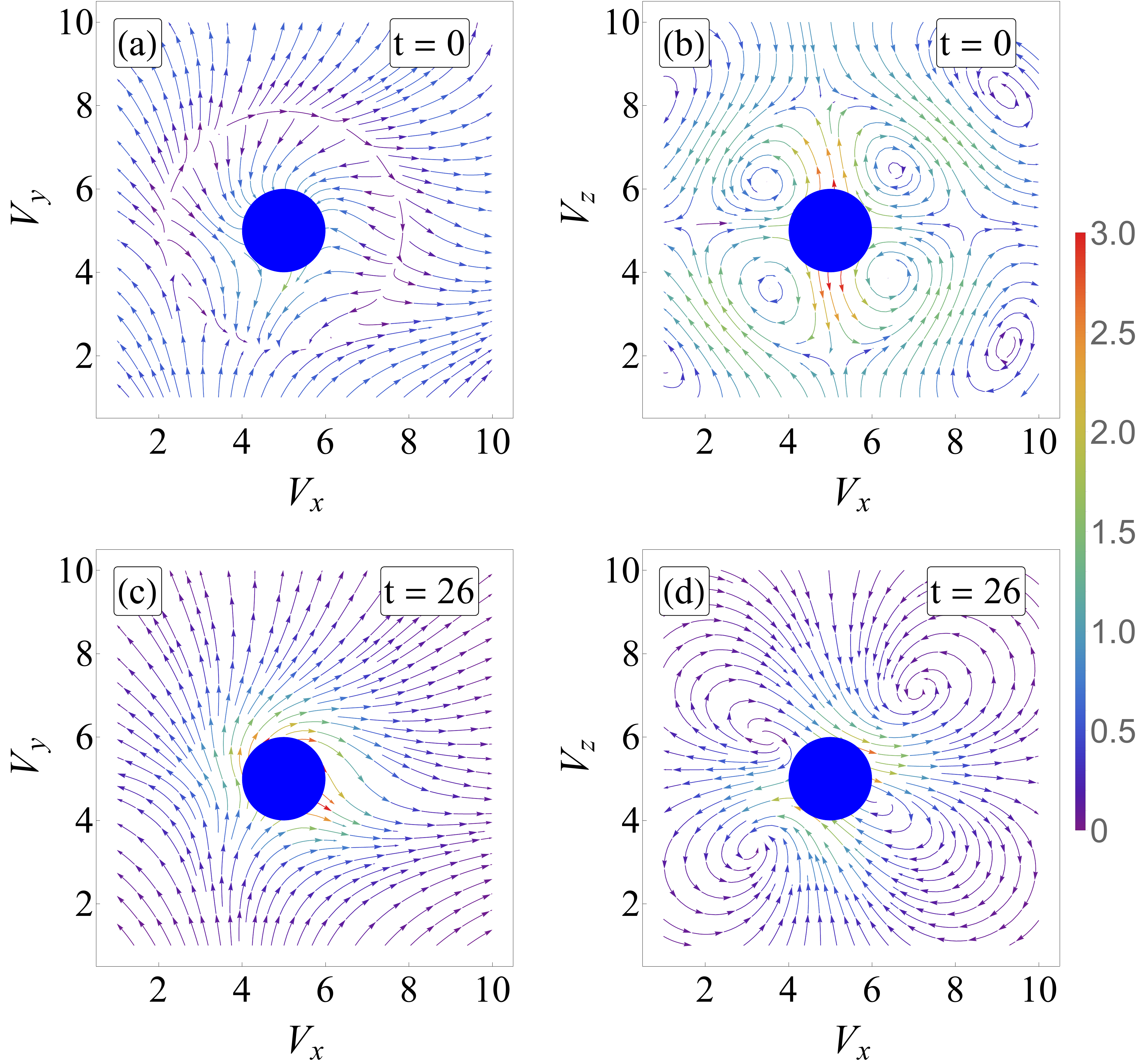}
\caption{(Color online) 
Real part of the flow field around an unsteady chiral swimmer in different planes at different times, $t = 0 \,((a), (b))$ and $t = 26 \,((c), (d))$.
The color bar indicates the strength of the velocity field.}
\label{Rot_stream}
\end{figure}

\section{Unsteady hydrodynamic flow}
\label{comp}
Fig.~\ref{vel_stream} depicts the comparison of the axisymmetric part of the velocity field obtained in the current model with that of Blake \textit{et al.} \citep{blake, magar} at times $t = 0, 26$. Note that the Blakes' solution is obtained in the quasi-steady limit of the Stokes equation. We consider the first three terms of the time dependent surface slip velocity given by Blake \textit{et al.}, 
\begin{subequations}
\begin{align}
B_1 &= B_{10} + B_{11}\cos \omega t + B_{12}\cos 2\omega t \, ,\\
B_2 &= B_{21}\sin \omega t + B_{22}\sin 2\omega t \, ,\\
B_3 &= B_{30} + B_{31}\cos \omega t + B_{32}\cos 2\omega t \, .
\end{align}
\label{eqn:bn}
\end{subequations}
 Here, $\omega$ is the time-dependent surface squirming frequency. We assume that the former frequency is the same as the frequency of the oscillatory motion of the unsteady chiral squirmer. While the first index of $B_{nl}(n=1,2,3; l = 0,1,2)$ on the RHS of Eq.~\ref{eqn:bn} implies the order of the mode amplitude, the second index is related to harmonics of a fundamental frequency of oscillation.
Also, in the Blakes' solution, the flow field is due to the force dipole, also known 
as hydrodynamic stresslet for $B_2(t) \ne 0$, and it decays as $1/r^2$. 
However, for $B_2(t) = 0$ the flow field is due to the source dipole, and it 
decays as $1/r^3$ (see Fig.~\ref{vel_stream}$(e)$ - $(h)$). 
Whereas, in our case, four strong vortices are formed in the vicinity of the swimmer due to the inertial effects, i.e., $Re_{osc} \gg Re_{trans}$. 
Thus, the flow field due to the oscillating stresslet ($o(1/r^2)$) shows dominating quadrupolar symmetry, see Fig.~\ref{vel_stream}(b),(d).
While the vortices near the body are stronger, away from the swimmer, they decay fast. 
The complex vorticity field is a common feature of an unsteady swimmer \citep{li} with a stagnation region. 
The stress exerted by the swimmer on the surrounding fluid causes the medium to strain. For an unsteady swimmer, the strain in the surrounding medium appears as vortices in the vicinity of the swimmer.
Besides, the flow field has stagnation points, where the fluid velocity is zero, connected by a circle (see Fig.~\ref{vel_stream}$(a)$). Similar behavior can be observed in the velocity field obtained by the Blakes' solution, at $t = 15.6$ (not shown in the figure). Since the swimmer here is unsteady, the appearance of the stagnation boundary is time-dependent, present at $t=0$ but absent at $t= 26$ (see figs.\ref{vel_stream}$(a)$,$(c)$). 
 Note that the radius of the stagnation region changes randomly with both time $t$ and the parameter $\beta$.

Fig.~\ref{Rot_stream}$(a)$-$(d)$ depicts the flow field of a chiral swimmer, i.e., considering the rotational part, at times $t = 0$ and $26$. At $t = 0$, the generated flow pattern comprises a whirl in place of the stagnation contour at a distance from the surface of the chiral swimmer due to the rotational flow (see Fig.~\ref{Rot_stream}(a)). 
However, as it is observed in Fig.~\ref{vel_stream}, after some time $( t = 26)$, this 
whirl disappears (see Fig.~\ref{Rot_stream}(c)). 
Note that the vortices are anti-clockwise (positive) for $t = 0$ and clockwise (negative) for $t = 26$ (see figs. \ref{Rot_stream}(b) and (d) and figs. \ref{vel_stream}$(b)$ and $(d)$). These vortices result from the flow field induced by the swimmer in earlier times. 
The direction of the vorticity depends on the nature of the swimmer \citep{chisholm}, i.e., puller or pusher, which oscillates with time 
(see Fig.~\ref{fig:slip_vel}).

\section{Response of an unsteady chiral swimmer to an external chemical gradient}
\label{sec:chem_grad}   

In general, microorganisms respond to the external gradient (stimulus) present in the environment by swimming towards \citep{kirchman} or away from the stimulus. The directional movement of the self-propelled body in an external chemical stimulus is known as chemotaxis.
 For instance, Paramecium shows chemotactic behavior with the help of receptors existing
either on its ciliary membrane \citep{doughty} or on the cell membrane of the organism 
\citep{oami}. 
In chemotaxis, the ligands bind to the receptors on the body's surface, which triggers the internal signaling network of the body. This network modifies the velocity and rotation rate of the swimmer so that it can follow up the chemical gradient \citep{bohmer,friedrichth,maity}. Therefore, the body adapts to the chemical stimulus and responds or relaxes accordingly.

 To understand the microswimmer's adaptation and relaxation mechanism in a chemical gradient, we adopt the Barkai-Leibler model \citep{barkai}, 
a simple model for information processing and memory via the internal biochemical network. 
This model has been widely used for bacterial and sperm chemotaxis \citep{friedrichth,bohmer}. 
The mechanism of adaptation and relaxation can be described by a simple 
dynamical system \citep{barkai},
\begin{subequations}
\begin{align}
\label{eq:adt,relax}
\sigma \, \dot{a_b} & = p_b (S_b + S) - a_b \ , \\
\mu \, \dot{p_b} & = p_b (1-a_b) \,,
\end{align}
\end{subequations} 
where $\sigma$ is the relaxation time, $\mu$ is the adaptation time, $a_b(t)$ is the output variable which controls the velocity and rotation rate of the swimmer, $p_b (t)$ is the dynamic sensitivity and $S_b(t)$ arises from the background activity in the absence of the ligand. The latter has a dimension of concentration. 
Note that these equations are applicable to weak concentration gradient limit only.
Under a constant stimulus $S(t) = S_c$, the system reaches a steady-state; the unsteady  swimmer propels along a fixed direction retaining its unperturbed swimming path. 
In the steady state, $a_b = 1$ and $p_b = 1/(S_b + S_c)$. 
Since there exist three time scales in the model, say, $\mu$, $\sigma$ and $2\pi/\omega$, 
 ($\omega$ is frequency of the oscillatory flow,
we have scaled two of them, i.e., $\mu$, $\sigma$, in terms of the third, i.e., $1/\omega$. 
The scaled ones read, $\tilde{\mu} = \mu \,\omega, \tilde{\sigma} = \sigma\,\omega$. 
However, for simplicity, we set $\tilde{\mu} = \tilde{\sigma} = 1$.

The chemotactic stimulus modifies the coefficients of the slip velocity (eqs.~\ref{slip_s}, ~\ref{slip_ex}), which are related to the velocity and the rotation rate of the body, in the following way, 
\begin{align}
X = X^{(0)} + X^{(1)}\,[a_b(t) - 1] \,,
\label{eq:beta_gamma}
\end{align}
where $X = (\delta_{l m}^{j, \prime k \, A}, \xi_{l m}^{j, \prime k \, C})$ and 
index $j$ takes values $0, 1$ and $k = 1$.
Here, $X^{0}$ terms are the unperturbed slip coefficients, i.e., in the absence of chemical gradient, whereas the terms with $X^{1}$ are the modified slip coefficients in the presence of the chemical gradient.
Since the velocity and the rotation rate solely dictate the path of the body, it is sufficient to consider only the first mode in the slip velocity (Eq.~\ref{slip}), i.e., we set $l = 1$. 
 As the slip coefficients enter in the velocity and rotation rate of the swimmer, we can take $X = (\bm{U}, \bm{\Omega})$ with $X^{(0)} = (\bm{U}^{(0)}, \bm{\Omega}^{(0)})$  and $X^{(1)} = (\bm{U}^{(1)}, \bm{\Omega}^{(1)})$.
 Note that while there is no external chemical gradient, $a_b (t)$ attains its steady-state value, which is one. Consequently, we can see that perturbation ($X^{(1)}$) in the slip coefficients vanishes in the absence of a chemical gradient, see Eq.~\ref{eq:beta_gamma}. 
The former equation is analogous to the variation of the curvature ($\kappa$) and torsion ($\tau$) of a sperm's trajectory in a chemical field \citep{friedrichth}. 
Note that $\kappa = |\bm{\Omega} \times \mathbf{U}|/|\mathbf{U}|^2$ and $\tau = |\bm{\Omega} \cdot \mathbf{U}|/|\mathbf{U}|^2$. 

\subsection{Radial chemical gradient}

The chiral swimmer is subjected to a radial chemical field, which is having the following form,
\begin{equation}
c(\mathbf{r}(t)) = \frac{c_r}{r(t)}\,,
\end{equation}
where $c_r$ is the constant which depends on the diffusivity of the ligands, 
i.e., the rate at which ligands are released from the chemical target and 
$r(t)$ is the distance between the center of mass position of the swimmer and the chemical source which is,
$r(t) = \sqrt{x(t)^2 + y(t)^2 + z(t)^2}$, if the source is placed at the origin $(0,0,0)$. 
The chemotactic stimulus \citep{friedrichth}, i.e., the local chemical concentration 
over the receptors reads, $S(t)= c(\mathbf{r}(t))$.
 As long as the chemical gradient is weak, there is no threshold for the swimmer to sense the chemical gradient. 
In the case of radial gradient, the relative change in stimulus level $\nabla S(t)/S(t) = 1/r(t)$ decides the instantaneous output $a_b(t)$ of the biochemical network, 
see Eqs.~\ref{eq:adt,relax}. 
This relative change is analogous to logarithmic (Weber's law) in the week gradient limit \citep{lazova, matt}. 

\begin{figure}[bh!]
\centering
\begin{center}
\includegraphics[scale = 0.15]{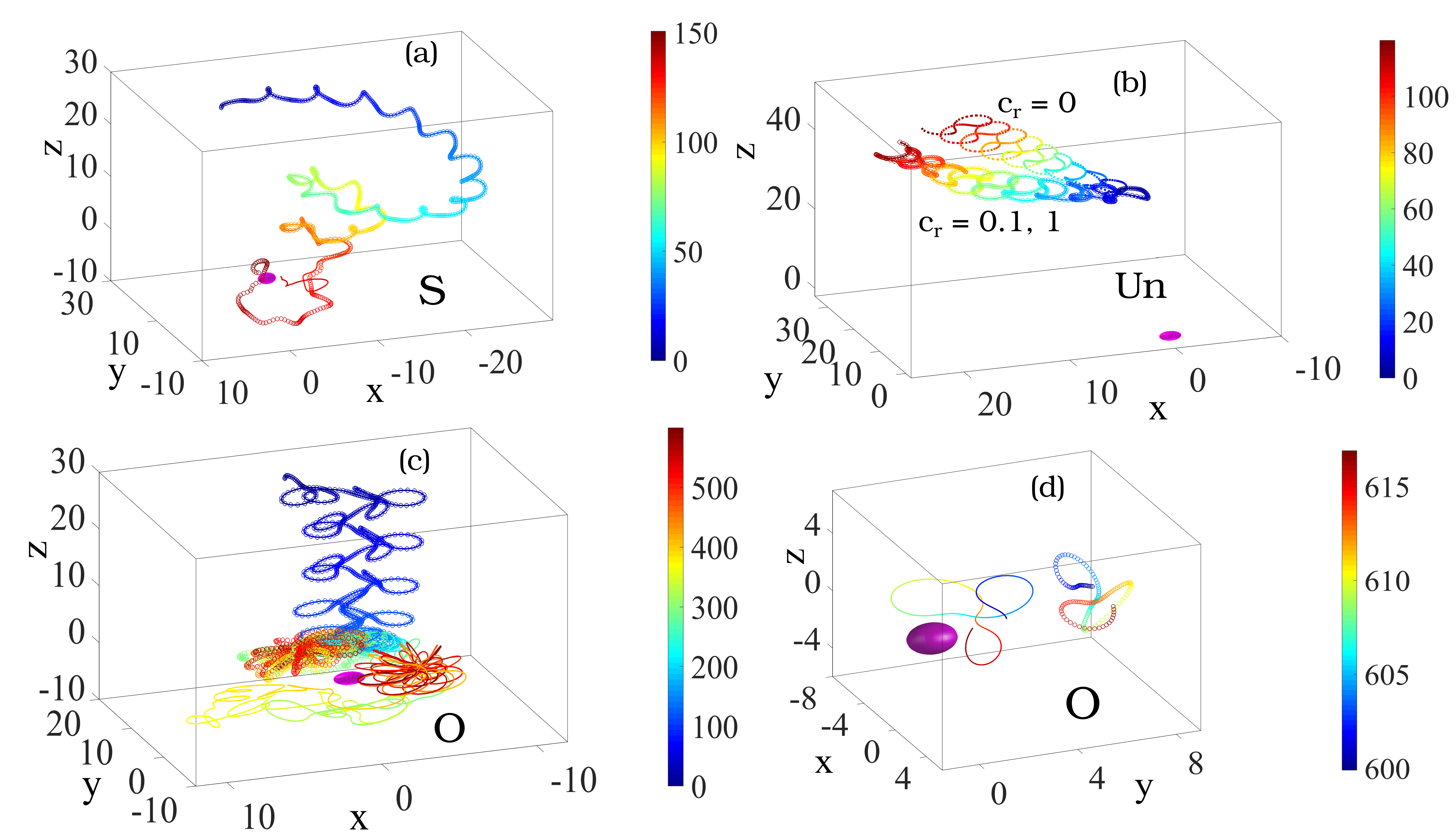}
\caption{(Color Online) 
Numerically obtained, the swimming paths of an unsteady chiral swimmer in a radial chemical gradient. 
 (a) depicts the successful chemotaxis (S) whereas $b$ corresponds to the unsuccessful chemotaxis (Un) and (c), (d) shows the trajectories corresponding to the orbiting state (O). 
The panel (d) illustrate a closer view to orbiting swimmer near the chemical target. 
The time along swimming paths are encoded by color shown in the color bar.
The source (sphere) is at $(0,0,0)$  and the swimmer's initial position is
$(0, 20, 25)$. Lines and circles correspond to $c_r = 0.1$ and $c_r = 1$, respectively. Dotted line in panel (b) corresponds to the $c_r = 0$ or unperturbed state. 
Here, the adaptation and relaxation times are set to $\tilde{\sigma} = \tilde{\mu} = 1$. 
The perturbed $(\bm{U}^{(0)}, \bm{\Omega}^{(0)})$  and 
and unperturbed $(\bm{U}^{(1)}, \bm{\Omega}^{(1)})$ 
quantities of $\bm{U}$ and $\bm{\Omega}$ are set as, 
$\bm{U}^{(0)}_0 = (0, 0, 2.39)$, 
$\bm{U}^{(1)}_0 = (0, 0, 0.239)$, 
$\bm{\Omega}^{(0)}_0 = (0.9, 0, 0.2)$, 
$\bm{\Omega}^{(1)}_0 = (-2, 0, 2)$,
$\bm{U}^{\prime\,(0)} = (0, 0, -2.3)$, 
$\bm{U}^{\prime\,(1)} = (0, 0, -0.23)$,
$\bm{\Omega}^{\prime\,(0)}_0 = (-0.5, 0, -0.5)$, 
$\bm{\Omega}^{\prime\,(1)}_0 = (-0.5, 0, -0.5)$,
$\bm{U}^{1\,(0)} = (0, 0, 1.4)$, and $\bm{U}^{1\,(1)} = (0, 0, 0.14)$. 
For (a) $\bm{\Omega}^{1 \, (0)} = (2 \sin(\pi/8), 0, -2 \cos(\pi/8))$ and 
$\bm{\Omega}^{1 \, (1)} = 10 \, \bm{\Omega}^{1 \, (0)}$.
For (b) $\bm{\Omega}^{1 \, (0)} = (\sin(\pi/8), 0, -\cos(\pi/8))$ and 
$\bm{\Omega}^{1 \, (1)} = 10 \, \bm{\Omega}^{1 \, (0)}$.
For (c),(d) $\bm{\Omega}^{1 \, (0)} = (\sin(11\pi/8), 0, -\cos(11\pi/8))$ and 
$\bm{\Omega}^{1 \, (1)} = 10 \, \bm{\Omega}^{1 \, (0)}$.}
\label{traj_rad}
\end{center}
\end{figure}

Our main assumptions are, $(i)$ the chemical concentration difference along the length of the body is irrelevant as the chemical signaling system receives a temporal stimulus, $(ii)$ the coefficients of the slip velocity are modified by the signaling system. 
Also, to have a minimal model, we can safely ignore the higher order terms corresponding to the oscillatory and transient 
parts, i.e., $j, \,k > 1$ in $\bm{U}(t)$ and $\bm{\Omega}(t)$.
Therefore, we are left with the following parameters, 
$\delta_{1 0}^{j,\prime k \, A(0)}, \xi_{1 0}^{j,\prime k \,C(0)}, \xi_{1 1}^{j,\prime k \, C(0)}$ for the unperturbed part, and 
$\delta_{1 0}^{j,\prime k \,A(1)}, \xi_{1 0}^{j, \prime k \, C(1)}, \xi_{1 1}^{j, \prime k \, C(1)}$ for the perturbed part with $j = 0, 1$ and $k = 1$. 
 The steady parts of the velocity $\bm{U_0}$ and the rotation rate $\bm{\Omega_0}$ are 
kept constant (see Eq.~\ref{eq:U} \& ~\ref{eq:W}). 
The transient parts of the velocity $(\bm{U}^{\prime 1})$ and rotation rate 
$(\bm{\Omega}^{\prime 1})$ of the swimmer decay very fast and may not play a significant role in the process of chemotaxis. Thus, we can keep them constant as well. 
Further, we can assume that the velocity of the swimmer is only slightly perturbed by the chemical gradient \citep{friedrichth}. For example, the ratio of perturbed to unperturbed slip coefficients $\delta_{1 0}^{j,\prime k \,A(1)}/\delta_{1 0}^{j,\prime k \, A(0)} \sim 0.1$. 
The former assumption is inspired by the chemotactic response of several natural microswimmers \citep{friedrichth, salek, pankratova}.
However, the rotation rate can be highly perturbed by the chemical gradient, i.e., when we apply a radial chemical gradient thus, we are considering 
$ \xi_{1 m}^{j \, C(1)} /\xi_{1 m}^{j \,C(0)} \sim 10 \, ( for\,  m = 0,1)$.

\begin{figure}
\centering
\includegraphics[scale= 0.55]{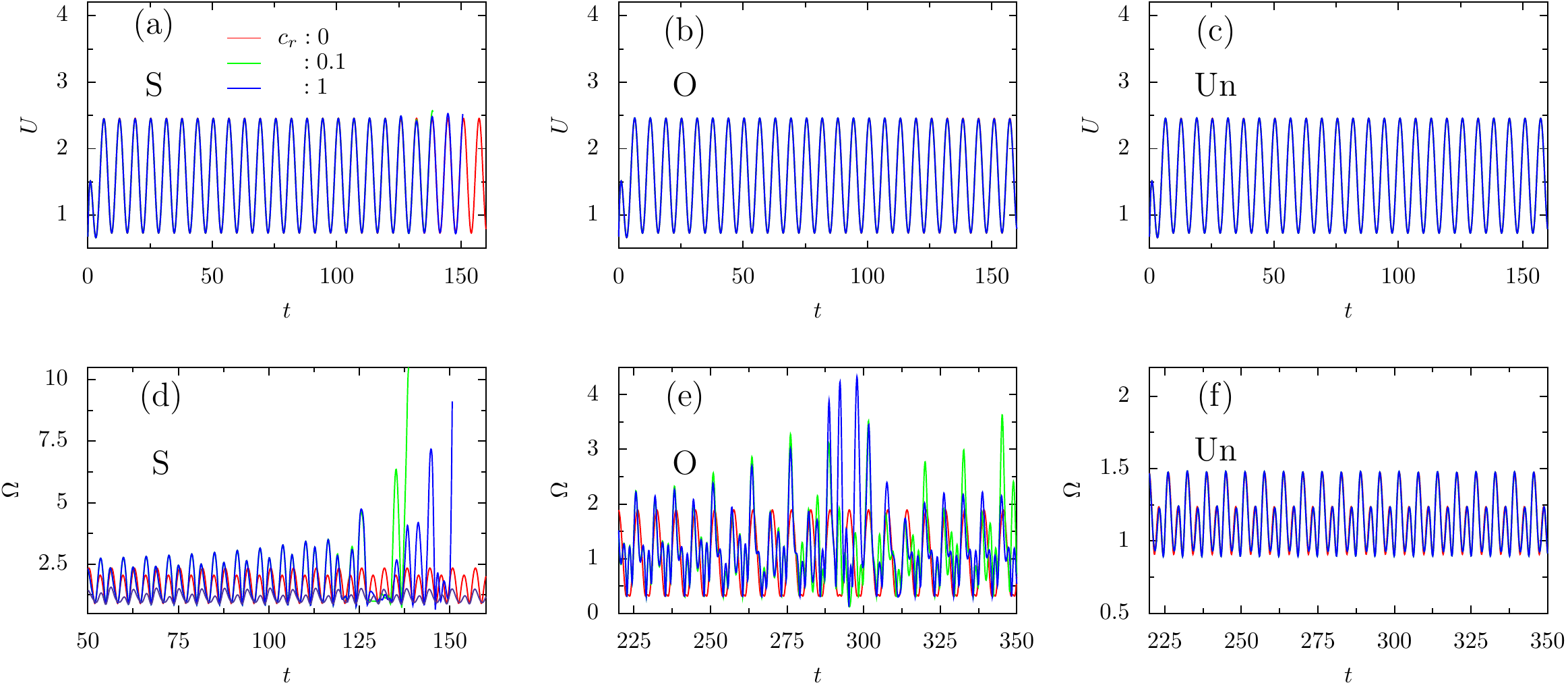}
\caption{(Color Online) Numerically obtained, 
the magnitudes of velocity $\bm{U}$ and rotation rate $\bm{\Omega}$ are plotted as a function of time for $c_r = 0, 0.1, 1$. 
$(a), (d)$ corresponds to successful chemotaxis, $(b), (e)$ belongs to orbiting swimmer, and $(c), (f)$ associated with unsuccessful chemotaxis. 
All other parameters are same as in Fig.~\ref{traj_rad}.}
\label{velom}
\end{figure}

We numerically solve the Eqs.~\ref{eqn:single}, with the modified velocity and rotation rate of the swimmer in the presence of 
the chemical gradient to obtain the trajectory of the swimmer.  
Fig.~\ref{traj_rad} depicts the chemotactic response of the swimmer for various strengths of the chemical gradient, where the source is placed at $(0, 0, 0)$. 
For the steady case, the swimmers' path is independent of $c_r$ \citep{maity}. However, for the present case, the trajectory of the unsteady swimmer depends on the strength of $c_r$. 
Depending on the strength of the slip coefficients, the swimmer exhibits 
successful chemotaxis (S), see Fig.~\ref{traj_rad}(a), unsuccessful chemotaxis (Un), 
see Fig.~\ref{traj_rad}(b), and an interesting closed orbit state (O) near the 
chemical source, see figs.~\ref{traj_rad}(c),(d).
In the case of successful chemotaxis, the swimmer moves in an irregular helical path, see Fig.~\ref{traj_rad}(a). 
We have observed that the chemotactic success rate is irregular for the values of $c_r$ in the range $O(10^{-1})$, and beyond this range, it is independent of $c_r$.  
Here, the swimmer's internal unsteadiness introduces the randomness in chemotactic navigation despite having a uniform gradient, giving rise to random motility or a persistent random walk. 
Therefore, $c_r$ plays a useful determinant of chemotactic efficiency. 
On the other hand, the rotation rate can also dictate the success of the chemotaxis.
For some values of the rotation rate, the chemical gradient can not profoundly influence the body to steer towards the chemical source leading to unsuccessful chemotaxis, see Fig.~\ref{traj_rad}(b). Despite having a little influence on the dynamic parameters of the swimmer in the former case, see Fig.~\ref{velom}(c),(f), the gradient effectively reroutes the swimmer, see Fig.~\ref{traj_rad}(b).
Additionally, as the orientation of the rotation rate changes swimmer exhibits 
the interesting bounded state where the swimmer moves in a closed orbit around the source of the chemical gradient. 
The latter is the transition state between successful and unsuccessful chemotaxis. 
In this state, the chemical gradient is strong enough to trap a swimmer in orbit but cannot influence it to reach the target. Thus, the orbiting state appears when there is a balance between the torque due to the radial chemical gradient and the torque generated by the swimmer. 
Note that the orbiting state has been reported previously for artificial microswimmers \citep{jinor,takagi}. 

Fig.~\ref{velom} shows the behavior of the velocity $U$ and the rotation rate $\Omega$ while the swimmer responds to the chemical gradient. 
The velocity of the swimmer is unchanged, whereas the rotation rate varies irregularly. 
Note that $\Omega$ diverges near the target in successful chemotaxis, see Fig.~\ref{velom}(d). 
It is due to the saturation of the internal chemotactic network responsible for the relaxation and adaptation mechanism. 
However, the mildly affected velocity remains the same near the target, see Fig.~\ref{velom}(a).  
In orbiting state, none of the dynamic parameters diverge as the swimmer never reaches the target, see Fig.~\ref{velom}(b)\&(e). 
Yet, similar to successful chemotaxis, the variation in rotation rate $(\Omega)$ is irregular but not in $U$. 
Consequently, the gradient can not strongly perturb the dynamic parameters in case of unsuccessful chemotaxis, see Fig.~\ref{velom}(c), (f).

\begin{figure}
\centering
\includegraphics[scale= 1.5]{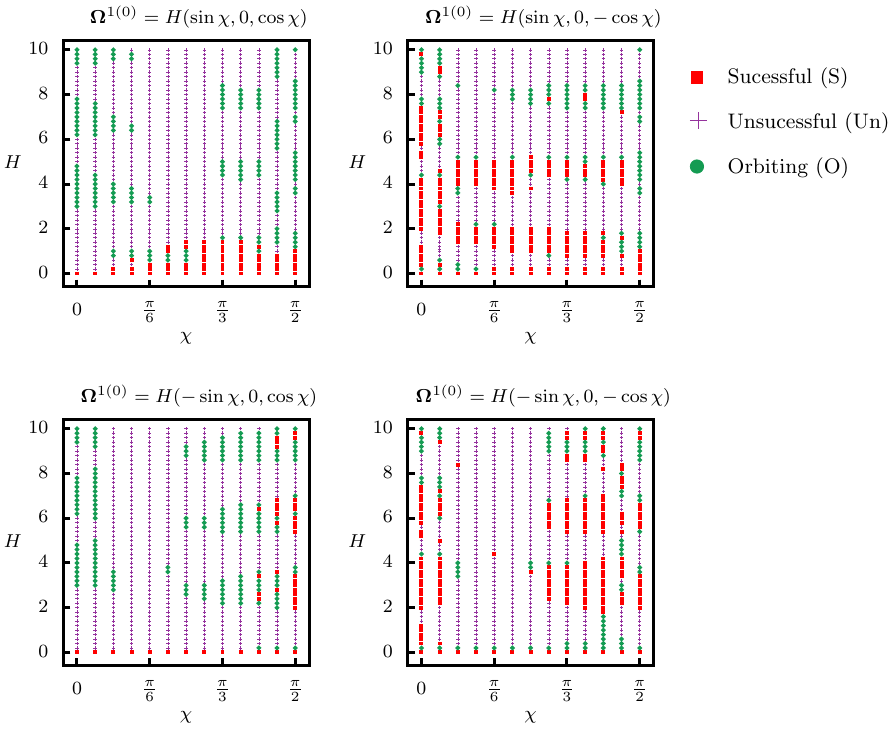}
\caption{(Color online) 
Numerically obtained, the state diagram depicting the swimming behavior of an unsteady chiral swimmer in a radial chemical gradient for various orientations and strengths of unperturbed oscillatory part of the rotation rate $\bm{\Omega}^{1 (0)}$ (see the text for more details). 
The rest of the parameters are same as those chosen in Fig.~\ref{traj_rad}.}
\label{fig:states}
\end{figure}

In the following, as the rotation rate plays a vital role in the chemotaxis of the chiral 
swimmer, we consider a situation where the strength of the chemical gradient is fixed 
$(c_r = 1)$, and vary the magnitude and orientation of the oscillatory part of the rotation rate 
$\bm{\Omega}^1$ systematically to understand the chemotactic behavior of the unsteady swimmer 
by keeping the oscillatory part of the velocity $\bm{U}^1$ constant. 
Note that, as mentioned earlier, the steady and transient parts of velocity and rotation rate, i.e., $\bm{U}_0, \bm{\Omega}_0, \bm{U}^\prime, \bm{\Omega}^\prime$, are 
kept constant (see the caption of Fig.~\ref{traj_rad}).   
We define the components of unperturbed $\bm{\Omega}^{1(0)}$ as 
$\xi_{11}^{1\, C(0)}(\pm)= \pm H \sin (\chi)$, 
$\xi_{11}^{1\, D(0)}(\pm)= 0$, 
$\xi_{10}^{1\, C(0)}(\pm)= \pm H \cos (\chi)$, 
where $\chi$ is the angle between the swimming direction $(\mathbf{t})$ and $\bm{\Omega}^{1(0)}$ 
and the magnitude of $\bm{\Omega}^{1(0)}$ is $H = \sqrt{\big(\xi_{11}^{1\, C(0)}(\pm)\big)^2 + \big(\xi_{10}^{1\, C(0)}(\pm)\big)^2}$.
For example, 
$\xi_{11}^{1\, C(0)}(+)= + H \sin (\chi)$, 
$\xi_{10}^{1\, C(0)}(+) = + H \cos (\chi)$, 
are along the $+ve$ $\bf n$ and $\bf t$ axes, respectively. 
Similarly, for the other combinations.
Note that, as mentioned earlier, due to external chemical gradient, the perturbed slip coefficients (see Eq.~\ref{eq:beta_gamma}) of the rotation rate varies as 
$\xi_{1 m}^{j \, C(1)} /\xi_{1 m}^{j \,C(0)} \sim 10$ ( for $m = 0,1$).

Fig.~\ref{fig:states} depicts the swimming behavior of an unsteady chiral swimmer in a radial chemical gradient for various orientations and strengths of unperturbed oscillatory part of the rotation rate $\bm{\Omega}^{1 (0)}$.
The successful chemotaxis is observed mostly for 
$\bm{\Omega}^{1(0)} = H (\sin\chi, 0, -\cos\chi)$ and 
$\bm{\Omega}^{1(0)} = H (-\sin\chi, 0, -\cos\chi)$. 
In the other combinations, the orbiting states appear. 
However, the unsuccessful states appear in all the combinations. 
For the case of a steady (time independent) chiral swimmer, i.e. for $H = 0$, chiral swimmer 
always exhibits successful chemotaxis \citep{maity}. 
Due to oscillatory nature of the swimmer, the swimming states 
show some retentiveness in the state diagrams (see Fig.~\ref{fig:states}).
The time-dependent part in the rotation rate $(\bm{\Omega}^1)$ introduces irregularness in the process of chemotaxis. 
As a result, higher magnitudes of $(\bm{\Omega}^1)$, i.e., $H$, reduce the success rate of chemotaxis, see Fig.~\ref{fig:states}. 
The varying strength of the oscillatory part of the rotation rate does not always lead to successful chemotaxis. 
Therefore, only those unsteady swimmers will successfully reach the chemical target, which will possess a specific rotation rate. 
For some combinations of $\chi$ and $H$, orbiting states appear between the successful and unsuccessful states. 
As it is mentioned earlier, the latter is generally the transition state between successful and unsuccessful chemotaxis.

\section{Conclusions}
\label{sec:conclusions}

In this work, we have investigated an unsteady chiral swimmer's behavior and its response to an external radial chemical gradient. We have prescribed a very general time-dependent slip velocity at the surface of the swimmer, and by using it, we have solved the unsteady Stokes equation for the velocity field. The corresponding time-dependent velocity and rotation rate of the swimmer are obtained in a simple form.
We showed that the unsteady inertial effects (added mass and Basset force) influence both the magnitude and the oscillatory behavior of the velocity, whereas they can only 
control the oscillatory behavior of the rotation rate. 
Thus, the velocity of the unsteady swimmer differs from the quasi-steady swimmer. 
Besides, an unsteady chiral swimmer produces vortices in the flow field. 
Indeed, the formation of vorticity causes a large amount of power dissipation. However, vortices may conceal mechanical strain fields that draw predators \citep{kiorboe}. 
Interestingly, due to time-dependent squirming, the swimmer oscillates between puller and pusher type. 

In the absence of a chemical gradient, the unsteady chiral swimmer moves in a deformed helical path with varying speeds. 
However, when the swimmer is placed in a chemical gradient, its slip coefficients are modified, and correspondingly the velocity and rotation rates are altered. 
We have observed that the rotation rate plays a vital role in the response of the swimmer to a radial chemical gradient. 
The rotation rate leads to randomness in the swimming path, which essentially controls the success rate of the chemotaxis. 
The chemotactic movement of the swimmer is a stochastic function of the chemoattractant diffusivity $(c_r)$ in the range $o(10^{-1})$ and independent of $c_r$ beyond that range. 

This study is helpful for understanding the behavior of unsteady ciliated microorganisms and their response to external gradients, such as chemical, temperature, etc. 
Also, it is helpful to design synthetic self-propelled bodies that can perform biomixing. 
 Note that in the case of chemically active phoretic particles, the effective slip velocity at the surface of the swimmer is due to the diffusiophoresis process \citep{Anderson, Saha, Kapral}. These artificial swimmers can sense the external chemical gradients by altering the slip velocity and exhibit chemotaxis. This process is analogous to what we have used in the current manuscript. However, the way slip velocity alters in phoretic particles is different from the present case.

\section*{Acknowledgments}
The authors are grateful for insightful discussions with G. P. Raja Sekhar. 
This work was supported by the Indian Institute of Technology Kharagpur, India.

\section*{Declaration of Interests}
The authors report no conflict of interest.

\appendix

\section{The non-dimensionalization of the unsteady Stokes equation}
\label{appendix_stre}

The Navier-Stokes equation is given by,
\begin{align}
\rho \bigg( \frac{\partial \mathbf{v}}{\partial t} + \mathbf{v} \cdot \nabla \mathbf{v}\bigg) & = \eta \nabla^2 \mathbf{v} - \nabla p \,,
\label{eq:NSt}
\end{align}
where 
$\mathbf{v}$ is the unsteady velocity field,
$p$ is the pressure field, 
$\eta$ is the viscosity of fluid, and 
$\rho$ is the density of the fluid. 
To make Eq.~\ref{eq:NSt} dimensionless, 
we use radius of the unsteady chiral swimmer $a$ as the length scale, 
$\tau = 1/\omega$ as the timescale, where $\omega$ is the frequency corresponding to 
the oscillatory flow of the swimmer, 
and $\eta U_0 /a$ as the scale for pressure. 
Furthermore, the velocity is scaled as $U_0$ (steady part of the swimmer's velocity). Thus we have two natural dimensionless numbers, $Re_{trans}$ (translational Reynolds number) and $Re_{osc}$ (oscillatory Reynolds number) defined as,
\begin{align}
Re_{trans} &= \frac{\rho U_0 a}{\eta} \, ,\\
Re_{osc} &= \frac{\rho \omega a^2}{\eta} \, .
\end{align}
Using the above scales the Navier-Stokes equation can be non-dimensionalized as,
\begin{align}
Re_{osc} \frac{\partial \mathbf{\tilde{v}}}{\partial \tilde{t}} + Re_{trans}\mathbf{\tilde{v}} \cdot \tilde{\nabla} \mathbf{\tilde{v}} &= \tilde{\nabla}^2 \mathbf{\tilde{v}} - \tilde{\nabla} \tilde{p} \, .
\end{align}
In the limit, $Re_{osc} \gg Re_{trans}$ the second term on the LHS vanishes. 
The Strouhal number $Sl$ is defined as,
\begin{align}
Sl = \frac{a \omega}{ U_0}\,.
\end{align}
Also, $Re_{osc}$ can be expressed as $Re_{osc} = Re_{trans} Sl$. 
Therefore, the momentum equation reduces to unsteady Stokes equation
\begin{align}
\label{eq:Dimless_US}
Sl Re_{trans} \frac{\partial \mathbf{\tilde{v}}}{\partial \tilde{t}} &= \tilde{\nabla}^2 \mathbf{\tilde{v}} - \tilde{\nabla} \tilde{p} \,.
\end{align}
The unsteady Stokes equation is scaled as $Sl Re$ (here $Re = Re_{trans}$). 
In our study, the velocity of the swimmer is determined from the slip velocity of the body with the aid of Lorentz reciprocal theorem. Surely, $Sl Re$ is absorbed into the slip coefficients or the coefficients of the flow field. 
Therefore, we do not write $Sl Re$ explicitly in the general solution of the flow field. We drop the tildes afterwards for better readability.

\section{Velocity field of an unsteady chiral swimmer}
\label{appendix}

To calculate the velocity field of an unsteady chiral swimmer moving in a stagnant fluid, 
we start with the general solution of the unsteady Stokes equation (Eq.~\ref{eq:Dimless_US}) which is given in dimensionless units as \citep{ padmavathi},
\begin{align}
\label{eq:Gen_sol}
\mathbf{v} = \sum^\infty_{l=1}\, \sum^l_{m=0}\, \sum^\infty_{n = 0}\, \Big(v_r \mathbf{e}_r + v_\theta \mathbf{e}_\theta + v_\phi \mathbf{e}_\phi \Big) \, e^{\lambda_n^2 t}\,, 
\end{align}
where the components read,
\begin{align*}
v_r & = l(l+1) \Big[{\cal{K}}^{n1}_{l m} r^{l-1} + \frac{{\cal{K}}^{n2}_{l m}}{r^{l+2}} + {\cal{K}}^{n3}_{l m} \frac{g_l(r\,\lambda_n )}{r} \Big]\, S_{l m}^n(\theta, \phi) \,, \\
v_\theta & = 
\Big[ (l+1) \,{\cal{K}}^{n1}_{l m} r^{l-1} - \frac{l \, {\cal{K}}^{n2}_{l m}}{r^{l+2}} 
 + {\cal{K}}^{n3}_{l m} \Big((l+1) \, \frac{g_l(r\,\lambda_n )}{r} 
   - \lambda_n \, g_{l+1}(r\,\lambda_n )\Big) \Big]\, \frac{\partial S_{l m}^n(\theta, \phi)}{\partial \theta} \nonumber \\
& + \Big[\frac{{\cal{K}}^{n4}_{l m} f_l(r\,\lambda_n ) + {\cal{K}}^{n5}_{l m} g_l(r\,\lambda_n )}{\sin\theta} \Big]\, 
\frac{\partial T_{l m}^n(\theta, \phi)}{\partial \phi} \, \\
v_\phi & = \frac{1}{\sin\theta} \Big[ (l+1) {\cal{K}}^{n1}_{l m} r^{l-1} - \frac{l {\cal{K}}^{n2}_{l m}}{r^{l+2}}
 {\cal{K}}^{n3}_{l m} \Big((l+1)\frac{g_l(r\,\lambda_n)}{r} - \lambda_n g_{l+1}
(r\,\lambda_n)\Big) \Big] \, \frac{\partial S_{l m}^n(\theta, \phi)}{\partial \phi} \nonumber \\
& - \Big[\frac{{\cal{K}}^{n4}_{l m} f_l(r\,\lambda_n) + {\cal{K}}^{n5}_{l m} \,
g_l(r\,\lambda_n)}{\sin\theta} \Big] \, \frac{\partial T_{l m}^n(\theta, \phi)}{\partial \theta} \,.
\label{eq:gen_comp}
\end{align*}
Here, 
$g_l(r \lambda_n) = \sqrt{\pi/(2\,r \lambda_n)} \, K_{l+\frac{1}{2}}$
and 
$f_l(r \lambda_n) = \sqrt{\pi/(2\,r \lambda_n)} \, I_{l+\frac{1}{2}}$
are the modified spherical Bessel functions of fractional order. 
$S_{l m}^n(\theta, \phi)$ and $T_{l m}^n(\theta, \phi)$ are the spherical harmonics defined in section \ref{sec:model}.

The boundary conditions at the surface of the chiral swimmer (at $r = a$) are prescribed
by the surface slip (Eq.~\ref{slip}).
However, in the far field, by setting $\mathbf{v}_{r \to \infty} = 0$, 
i.e., solving the velocity field in the laboratory frame of reference for $l = 1$ mode we get,
\begin{align*} 
{\cal{K}}_{1 m}^{n1} &  = -\frac{U_n}{2}\, , \\
{\cal{K}}_{1 m}^{n2} &  = \frac{-a^3 (-2\delta_{1m}^n g_1(\lambda_n a) + \lambda_n a\, g_2(\lambda_n a)U_n)}{2(3g_1(\lambda_n a)- \lambda_n a \, g_2(\lambda_n a))}\, , \\
{\cal{K}}_{1 m}^{n3} &  = \frac{a (-2\delta_{1m}^n + 3U_n)}{2(3g_1(\lambda_n a)- \lambda_n a \, g_2(\lambda_n a))}\, , \\
{\cal{K}}_{1 m}^{n4} &  = 0\, , \\
{\cal{K}}_{1 m}^{n5} &  = \frac{\Omega_n a - \xi_{1m}^n}{g_1(\lambda_n a)}\,.
\end{align*}
For  $l > 1$ modes,
\begin{align*} 
{\cal{K}}_{l m}^{n1} &  = 0\, , \\
{\cal{K}}_{l m}^{n2} &  =  \frac{- a^{l+2}\delta_{lm}^n g_l(\lambda_na	)}
{\lambda_na \, g_{l+1}(\lambda_na) - g_l(\lambda_na) (2l +1)}\, , \\
{\cal{K}}_{l m}^{n3} &  = \frac{a\delta_{lm}^n}{\lambda_na \, g_{l+1}(\lambda_na) -g_l(\lambda_na)(2l+1)}\, , \\
{\cal{K}}_{l m}^{n4} &  = 0\, , \\
{\cal{K}}_{l m}^{n5} &  = \frac{ -\xi_{lm}^n}{g_l(\lambda_na)}\,.
\end{align*}
Note that, $\lambda_n^2 = i\, n$ $(i = \sqrt{-1})$ stands for oscillatory flow while $\lambda_n^2 = - n $ for transient (decaying) flow. 
In Fig.~\ref{vel_stream}, we have plotted the real part of the general velocity field given above. 
In the main text, $n = j \, (\lambda_j^2 = i j)$ and $n = k \,(\lambda_k^2 = -k)$
correspond to oscillatory and transient part of the flow, respectively. 
Accordingly, we have denoted the slip coefficients and the spherical harmonics without 
prime symbol for the oscillatory part and with prime symbol for the transient part.

\section{Derivation of velocity and rotation rate of the unsteady swimmer using Lorentz reciprocal theorem}
\label{appendix2}

Let $\mathbf{v}_1$ and $\mathbf{v}_2$ both satisfies unsteady Stokes equation. The corresponding stress tensors are $\bm{\sigma}_1$ and $\bm{\sigma}_2$ respectively. Notably, $\nabla \cdot \bm{\sigma}_i = \partial \mathbf{v}_i/\partial t$ with $i = 1,2$. We know that \citep{kim},
\begin{align}
\mathbf{v}_1 \cdot (\nabla \cdot \bm{\sigma}_2) - \mathbf{v}_2 \cdot (\nabla \cdot \bm{\sigma}_1) &= \nabla \cdot (\mathbf{v}_1 \cdot \bm{\sigma}_2 - \mathbf{v}_2 \cdot \bm{\sigma}_1) \,.
\end{align}
By integrating over the volume $V$ on both sides in the above equation, and then using the Gauss divergence theorem on RHS, we can obtain the Lorentz reciprocal theorem \citep{kim}.
It reads,
\begin{align}
\varointclockwise_S \mathbf{v}_1 \cdot (\bm{\sigma}_2 \cdot \mathbf{n})dS - \int_V \mathbf{v}_1 \cdot (\nabla \cdot \bm{\sigma}_2) dV &= \varointclockwise_S \mathbf{v}_2 \cdot (\bm{\sigma}_1 \cdot \mathbf{n})dS - \int_V \mathbf{v}_2 \cdot (\nabla \cdot \bm{\sigma}_1) dV \, .
\label{lorentz}
\end{align}
Assuming similar time dependency of both the solutions we can write 
$\mathbf{v}_i = \mathbf{v}_i(r,\theta, \phi)e^{\lambda^2 t}$ \citep{padmavathi} with $i=1,2$. Thus, Eq.~(\ref{lorentz}) reduces to
\begin{align}
\varointclockwise_S \mathbf{v}_1 \cdot (\bm{\sigma}_2 \cdot \mathbf{n})dS  &= \varointclockwise_S \mathbf{v}_2 \cdot (\bm{\sigma}_1 \cdot \mathbf{n})dS  \,.
\label{lorentz1}
\end{align}
The solution $(\mathbf{v}_1, \bm{\sigma}_1)$ is due to a force-free and torque-free unsteady swimmer. On the other hand, $(\mathbf{v}_2, \bm{\sigma}_2)$ is a solution due to a identical passive body subjected to a force $(\mathbf{F}_2)$ and torque $(\mathbf{T}_2)$ \citep{padmavathi}. 
Thus, we can write \citep{stone}
\begin{align}
\mathbf{F}_2 \cdot \mathbf{U}(t) &= - \int_{S} \mathbf{n}\cdot \bm{\sigma}_2\cdot \mathbfcal{V}_s dS \,, \\
\mathbf{T}_2 \cdot \bm{\Omega}(t) &= - \int_{S} \mathbf{n}\cdot \bm{\sigma}_2\cdot \mathbfcal{V}_s dS \,.
\end{align}
For an unsteady swimmer moving with translational velocity $\mathbf{U}(t)$ and with rotation 
rate $\bm{\Omega}(t)$, the flow field on its surface reads, $\mathbf{v}_1(a) = \mathbf{U}(t)+ \bm{\Omega}(t) \times (a \mathbf{e}_r)+ \mathbfcal{V}_s(\theta, \phi, t)$. 
Here, $\mathbf{e}_r$ is the unit outward normal on the surface of the swimmer and $a$ is its radius.  Also, $\mathbfcal{V}_s$ is the time dependent surface slip velocity, see section \ref{sec:model}. Hence, the velocity and rotation rate are read as,
\begin{align}
\label{eq:Uapp}
\mathbf{U}(t) & = \mathbf{U}_0 
+ \sum_{j=1}^{\infty}  \Re \left[ \mathbf{U}^j \, e^{\lambda^2_j t} \right] + \sum_{k=1}^{\infty}  \Re \left[ \mathbf{U}^{\prime k} \, e^{\lambda^2_k t} \right]\,,\\
\label{eq:Wapp}
\bm{\Omega}(t) & = \bm{\Omega}_0 + 
\sum_{j=1}^{\infty}  \Re \left[ \bm{\Omega}^j \, e^{\lambda^2_j t} \right] + 
\sum_{k=1}^{\infty}  \Re \left[ \bm{\Omega}^{\prime k} \, e^{\lambda^2_k t} \right]\,,
\end{align}
where, the first term on the RHS is corresponding to the steady (time independent) velocity, the second term is related to oscillatory part of the velocity and the third term is related to the transient (decaying) part of the velocity. 
Here,  
\begin{align}
 \label{eq:U_0app}
 \mathbf{U}_0 & = \frac{2}{3}\, \left(\delta_{1 1}^{0A}\,\mathbf{n} + \delta_{1 1}^{0B}\,\mathbf{b} + \delta_{1 0}^{0 A}\,\mathbf{t}\right) \,,\\
 \label{eq:U_japp}
\mathbf{U}^{j, \prime k} & = \frac{2}{3} \,
\frac{(1+ a \lambda_{j,k})}{\left(1 + a \lambda_{j,k} + \frac{a^2 \lambda_{j,k}^2}{3}\right)} \, \left(\delta_{1 1}^{j, \prime k \, A}\,\mathbf{n} + \delta_{1 1}^{j,\prime k \, B}\,\mathbf{b}+ \delta^{j,\prime k \, A}_{1 0}\,\mathbf{t}\right)\,, \\
 \label{eq:Omega_0app}
\bm{\Omega}_0 & = \frac{\xi_{1 1}^{0C}}{a} \mathbf{n} + \frac{\xi_{1 1}^{0D}}{a}  \mathbf{b} +  \frac{\xi^{0 C}_{1 0}}{a} \mathbf{t} \, ,\\
 \label{eq:Omega_japp}
\bm{\Omega}^{j, \prime k} & = \frac{\xi_{1 1}^{j,\prime k \, C}}{a} \mathbf{n} + \frac{\xi_{1 1}^{j,\prime k \, D}}{a}  \mathbf{b} +  \frac{\xi^{j,\prime k \, C}_{1 0}}{a} \mathbf{t}  \,.
\end{align}
The symbols have been discussed in section \ref{sec:model}. As the Basset (history) and added mass forces act on the unsteady swimmer, the force-free condition contains the active force, hydrodynamic drag force, and unsteady inertial force. 
Furthermore, the effective velocity of the body, see Eqs.~(\ref{eq:U_0app}), ~(\ref{eq:U_japp}), considers all the former forces. 
Thus, the first equation of Eqs.~(\ref{eqn:single}) can be found. 
Since the unsteady inertia does not influence the rotation rate of the swimmer  
(see Eqs.~(\ref{eq:Omega_0app}), ~(\ref{eq:Omega_japp})), 
the torque-free condition consists of active torque and hydrodynamic torque acting on the swimmer. Consequently, we can determine the remaining three equations of Eqs.~(\ref{eqn:single}) \citep{kim}.


\bibliographystyle{jfm}

\end{document}